\documentclass[journal]{IEEEtran}
\usepackage{graphicx}
\usepackage{url}
\usepackage{algorithm,algorithmic}
\usepackage{multirow}
\usepackage{color}

\usepackage{tabularx}

\usepackage{cite}

\ifCLASSINFOpdf
\else
\fi
\usepackage{amsmath}
\usepackage{amsfonts}
\begin{document}
%
\title{Semantic-Preserving Linguistic Steganography by Pivot Translation and Semantic-Aware Bins Coding}
%
%
%

\author{Tianyu Yang, Hanzhou Wu, \emph{Member, IEEE}, Biao Yi, Guorui Feng and Xinpeng Zhang
\thanks{\emph{Corresponding author: Dr. Hanzhou Wu, E-mail: h.wu.phd@ieee.org}}
}

%
%

\markboth{}%
{}
%



\maketitle

\begin{abstract}
Linguistic steganography (LS) aims to embed secret information into a highly encoded text for covert communication. It can be roughly divided to two main categories, i.e., modification based LS (MLS) and generation based LS (GLS). Unlike MLS that hides secret data by slightly modifying a given text without impairing the meaning of the text, GLS uses a trained language model to directly generate a text carrying secret data. A common disadvantage for MLS methods is that the embedding payload is very low, whose return is well preserving the semantic quality of the text. In contrast, GLS allows the data hider to embed a high payload, which has to pay the high price of uncontrollable semantics. In this paper, we propose a novel LS method to modify a given text by pivoting it between two different languages and embed secret data by applying a GLS-like information encoding strategy. Our purpose is to alter the expression of the given text, enabling a high payload to be embedded while keeping the semantic information unchanged. Experimental results have shown that the proposed work not only achieves a high embedding payload, but also shows superior performance in maintaining the semantic consistency and resisting linguistic steganalysis. 
\end{abstract}


\IEEEpeerreviewmaketitle

\section{Introduction}
\IEEEPARstart{s}{teganography} \cite{Simmons:paper} is referred to as the art of embedding secret information in a cover such as digital image without significantly distorting the cover. The resulting stego containing hidden information does not introduce noticeable artifacts and will be sent to a receiver for information extraction via an insecure channel such as the Internet. The purpose of steganography is to conceal the existence of the present secret communication so that secret information can be reliably conveyed to the receiver without arousing suspicion from the monitor. Steganography is not to replace cryptography but rather to enhance the security using its concealment, promoting it to become quite important in modern information security. 

The carrier used for steganography can be any media object such as image \cite{Wu:2017} and video \cite{Chen:2021}. Among these media objects, text is very suitable for steganographic activity because text is the most commonly used information-carrier in our daily life. Especially, the rapid development of mobile social networks enables people to easily realize \emph{text steganography}, which is also called \emph{linguistic steganography (LS)} or \emph{natural language steganography (NLS)}. On the one hand, it is very convenient for the steganographer to integrate into the social network via a mobile terminal. On the other hand, the seemingly-normal stego texts can be easily overshadowed by the huge number of ordinary texts. In other words, LS can be easily concealed by the huge number of normal activities over social networks. Furthermore, for a data receiver, he can keep silent, collect the stego texts and further extract the hidden information without taking any suspicious interaction with anyone else. It indicates that LS (over social networks) even conceals the real receiver. 

However, unlike many other media objects that have a large redundant space (e.g., digital images), texts are often highly encoded and thus show low redundancy characteristics. Hiding additional information into texts is more challenging. In recent years, with the rapid development of deep learning and natural language processing, increasing works are proposed to realize LS, which can be roughly divided into two major categories that are modification based LS (MLS) and generation based LS (GLS). MLS embeds secret information into a given text by slightly modifying the text such as synonym substitution \cite{KeithWinstein:paper, Huo:paper},  typo confusion \cite{Topkara:paper} and syntactic transformation \cite{Kim:paper}. In brief summary, MLS requires a cover text in advance. Instead, GLS skips the cover text and directly produces a text carrying secret information for steganography such as \cite{Guo:paper, Kang:paper, Fang:paper, Ziegler:paper, Yang:paper, Yang:paper2}.

To be specific, mainstream GLS methods adopt a good language model to generate stego texts, which can be generalized by two important steps. First, a language model is well trained on a large-scale corpus so that the trained language model can automatically generate texts with high semantic quality. Then, an information encoding method is used during text generation so that a sequence of words (carrying secret information) can be generated to form the stego text using the trained language model. A key problem is how to encode the secret information, i.e., how to build the mapping relationship between words and secret bits. Many methods are carried out around this problem. For example, Fang \emph{et al.} \cite{Fang:paper} divide a vocabulary into multiple subsets and select the token with the highest probability in the subset corresponding to the
secret binary stream as the output. Yang \emph{et al}. \cite{Yang:paper} propose two novel techniques called fix-length coding (FLC) and variable-length coding (VLC) according to the probability distribution of candidate words to map secret bits to words. Because GLS does not require a given text, the size of the payload is often higher than MLS.

In recent years, much attention has been paid to narrowing the difference in statistical distribution between stego texts and natural texts \cite{Ziegler:paper, Cai:paper}. However, for these methods, the term ``imperceptibility'' refers to language fluency as well as statistical similarity, rather than semantic consistency. Yang \emph{et al}. \cite{Yang:paper3, Yang:paper4} propose novel strategies to make the semantic expression conform to the context, yet expressing the accurate semantics remains impossible, making it difficult to adapt to realistic scenarios. Therefore, though GLS has the higher payload, semantic consistency is still a very challenging problem.

On the other hand, MLS can be further divided into different categories. For example, word or phrase level MLS methods \cite{KeithWinstein:paper, Huo:paper} alter part of the cover text to embed secret information. These methods often first build a synonym dictionary to find the candidates, and then encode each candidate with various techniques such as binary tree \cite{Chiang:paper} and Huffman coding \cite{Bergmair:paper}. However, word or phrase level MLS arts may cause ambiguity because of partial replacement, and they suffer the problem of low embedding payload. Sentence level MLS tends to convert a sentence into another form maintaining the same meaning, e.g., word ordering \cite{Chang:paper} and syntactic analysis \cite{Murphy:paper}. Sentence level MLS modifies the entire sentence rather than part. So, it can maintain the semantic coherence of the entire sentence, which, however, should pay the high price of lower embedding payload and can be detected by deep learning tools \cite{Yi:paper, Yi:paper2}. 

Based on the aforementioned analysis, we propose a novel sentence level MLS algorithm using a GLS-like information encoding technique in this paper, which not only well controls semantic consistency, but also ensures a high steganographic payload. The proposed method pivots the original cover text between two different languages (i.e., English and German) to address the semantic inconsistency problem in GLS and the low embedding payload problem in MLS. In detail, we train a sequence-to-sequence-to-sequence (Seq2Seq2Seq) model to achieve the transformation of English-to-German-to-English (En2Ge2En), and embed secret information in the decoding stage of the second sequence-to-sequence (Seq2Seq) model, i.e., German-to-English (Ge2En) model. We regard the German text as the unchanged semantic information of the cover text, and only change the English expression during pivoting. Moreover, we propose a novel method called Semantic-aware Bins (SaBins) to realize information encoding. As a result, the stego text (in English) is different from the cover text but has the similar meaning. Due to semantic consistency and the fact that the receiver does not need the original text, the proposed method is quite suitable for practice, which significantly outperforms mainstream GLS methods that require the receiver to hold the trained language model, e.g., \cite{Kang:paper, Yang:paper}.

The main contributions of this paper are as follows:

\begin{itemize}
	\item By altering the expression of a given text through pivot translation and semantic-aware bins coding, the proposed work achieves a significantly higher embedding payload while well controlling the semantic information compared with many MLS methods, bringing MLS closer to GLS.
	\item Unlike GLS that uses a specific corpus to train a language model to be pre-shared and thus has limited usage scenarios, by applying a self-supervised training strategy for paraphrase generation, texts with any style can be used for semantic-preserving LS without model pre-sharing in our work, which builds a bridge between MLS and GLS.
	\item Extensive experiments have indicated that, compared with related works, the proposed work not only achieves a high embedding payload, but also shows superior performance in preserving the semantics and resisting steganalysis.
\end{itemize}

The rest structure of this paper is organized as follows. We first present preliminaries in Section II. Then, we introduce the proposed work in detail in Section III, followed by convinced experimental results and analysis in Section IV. We conclude this work in Section V.

\section{Preliminary Concepts}
In this section, we briefly introduce related concepts so that we can better introduce the proposed method in Section III.

\subsection{Sequence-to-Sequence (Seq2Seq) Model}
As introduced in \cite{seq2seq:paper}, a Seq2Seq model maps a sequence to another sequence where the length of the input sequence and the length of the output sequence may be different from each other. A Seq2Seq model generally consists of an encoder and a decoder. The encoder encodes the input sequence $\textbf{x} = \{x_1$, $x_2$, $...$, $x_L\}$ (whose length is $L>0$) into a hidden vector. The decoder then calculates the output vector $\textbf{y}_t$ = $\{y_{t,1}, y_{t,2},$ ..., $y_{t,N}\}$ at time $t$ based on the hidden vector and the previous $t-1$ output vectors $\textbf{y}_{1\sim (t-1)}$, where $N$ equals the number of all possible values of each element to be generated, e.g., $N$ is the number of all possible tokens (i.e., the size of the vocabulary) for generating a word sequence. By applying softmax to $\textbf{y}_t$, we can generate the present element by, for example, choosing the element corresponding to the largest prediction probability as the output. By collecting all generated elements, we are able to construct the final output sequence. A Seq2Seq model can be realized by using the long short-term memory (LSTM) \cite{lstm:paper} architecture or other cells. We refer a reader to \cite{seq2seq:paper} for details. 

\subsection{Natural Language Translation (NLT)}
Translation is a most basic task in natural language processing (NLP). The purpose of translation is to convert a text in one language to another with the same meaning. Traditional Statistical Machine Translation (SMT) systems \cite{Osborne:book, Brown:paper} are based on statistical models derived from a good corpus, well-designed linguistic rules, or a combination of them. With the rapid development
of deep learning, most machine translation systems in use today are Neural Machine Translation (NMT) systems \cite{Cho:paper, Luong:paper}. In other words, translation can be treated as a Seq2Seq problem and realized by using a Seq2Seq model.

\begin{figure*}[!t]
\centering
\includegraphics[width=\linewidth]{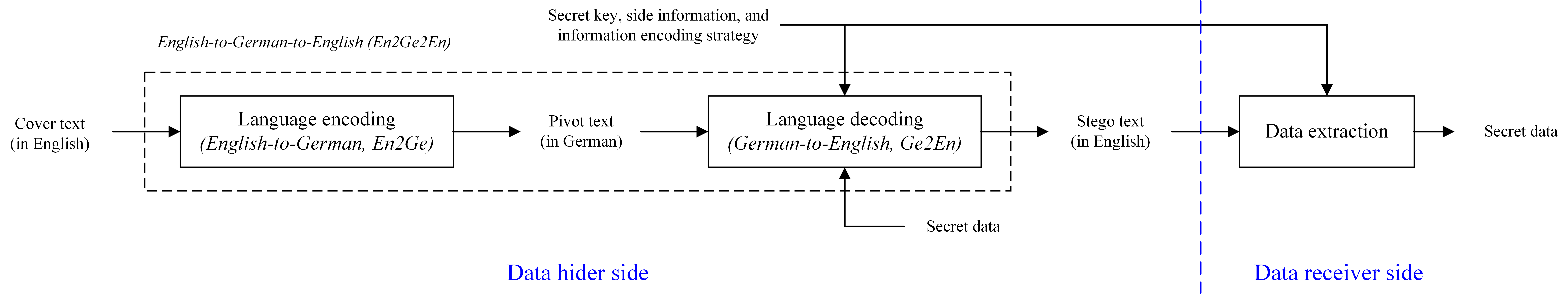}
\caption{Sketch for the proposed semantic-preserving linguistic steganographic framework based on pivot translation and semantic-aware bins coding.}
\end{figure*}

\subsection{Byte Pair Encoding (BPE) and Subword}
For text generation tasks, it is impossible to fill the vocabulary with all words because there are countless English words in the world and training a good text generation model will be extremely time consuming if the vocabulary is too large. However, if we use a small vocabulary, there will be many words that are not in the vocabulary and cannot be represented. Meanwhile, although 26 letters are enough to represent all English words, character-level tokenizer will result in too small granularity and make the training of a deep model too difficult.

Byte pair encoding (BPE) and subword \cite{subword:paper} are exploited to solve the above out-of-vocabulary (OOV) problem. It divides data into consecutive byte pairs and merges byte pairs according to the frequency in the training data, and finally forms a subword vocabulary. Subword is corresponding to a word segmentation method between character level and word level. A subword vocabulary consists of a large number of subwords and complete words. In recent research, BPE and subword have become one of the core technologies in NLP. In this paper, unless otherwise specified, we use the subword strategy based on BPE \cite{subword:paper} for the proposed method by default.

\subsection{Transformer}
Transformer \cite{transformer:paper} has been widely used in the field of NLP in recent years due to the superiority in feature extraction and expression. Compared to previous Seq2Seq models \cite{lstm:paper, seq2seq:paper1, Chung:2014}, Transformer lies in parallel computing, which greatly reduces the time-consuming problem of Seq2Seq model training. The introduction of the position vector also prevents the loss of the relative position information between tokens.

Furthermore, thanks to attention mechanism \cite{attention:paper}, Transformer can automatically learn the correlation between words
in the entire text. On the basis of attention mechanism, multi-head attention mechanism is introduced to help the model focus on several aspects of information and synthesize at the last step, which helps the network capture richer features.

\subsection{Paraphrase Generation}
Paraphrase is a method to explain the meaning of a given text in another expression using the same language. We can regard the semantic information of the paraphrase the same as the original text. Paraphrase generation is also a method of data argument in NLP, and has been used for performance improvement in several applications, such as information extraction \cite{paraphrase:paper} or question answering \cite{QA:paper}.

Paraphrase generation is also related to linguistic steganography. Generating semantic-consistent text is a jointly goal of paraphrase and MLS. As a generation task, paraphrase can be the bridge between MLS and GLS. Considering using a specific dataset will lead to limited usage scenarios such as asking questions \cite{QA:paper2} or picture descriptions \cite{coco:paper}, we exploit a self-supervised method of generating paraphrases instead in this paper, so that any text can be used as a cover in our proposed method. We will show the details in Section III. 

\subsection{Bins Coding}
Fang \emph{et al}. \cite{Fang:paper} propose a novel GLS method that randomly partitions the vocabulary with a fixed size into $2^b$ bins evenly in advance, where $b$ is a pre-determined integer. Each token in the vocabulary belongs to exactly one bin. At each step, they select one bin based on the secret string. Then, they select the token with the highest conditional probability from the chosen bin. As a result, it embeds $b$ bits at each generation step. Notice that, throughout this paper, unless mentioned, we ignore the common-token variant introduced in \cite{Fang:paper}.

The advantage of this novel method is that the data receiver neither needs to obtain the conditional probability, nor does he need the language model. Although there are many improved GLS methods afterwards such as \cite{Guo:paper, Kang:paper, Yang:paper, Yang:paper2}, they are subject to the constraint that the receiver has to calculate the conditional probability distribution for each token with the original model.

\subsection{Synonym Substitution}
The semantics of synonyms are similar to each other so that synonym substitution is a common method in MLS. It replaces the words in the cover text with synonymous words based on a dictionary and an information encoding strategy. For example, the two words ``see'' and ``look'' are corresponding to ``0'' and ``1'', respectively. The word ``see'' can be used to replace the word ``look'' in a cover text in order to hide the bit ``0''. 

To prevent ambiguity, when using synonym substitution, a large natural corpus is used as supplement, such as Google n-gram data \cite{googleNgram:paper} to evaluate the frequency of n-grams before and after the modified position in the large natural corpus. If the frequency exceeds a threshold \cite{ChangLS:paper} or is close to the original text \cite{Chang:paper}, the replacement is considered reasonable.

\section{Proposed Method}
\subsection{General Framework}
In order to provide a high embedding payload and keep the semantics unchanged, we propose to embed secret information into a given text by pivot translation and semantic-aware bins coding. As shown in Fig. 1, the proposed framework includes three phases, i.e., language encoding, language decoding and data extraction. The data hider will perform language encoding and language decoding, and the data receiver will perform data extraction. The goal of language encoding is to encode a given cover text in English into a new text in German (defined as pivot text). The two texts have different expressions but have the same meaning. During language decoding, the pivot text will be decoded into a stego text in English, whose semantic information is quite close to the original cover text as well as the pivot text. The stego text carries secret information and will be sent to the data receiver, who will try to reconstruct the secret information from the stego text according to the secret key and side information. Below, we show more details. 

\subsection{English-to-German-to-English (En2Ge2En) Model}
We propose to use paraphrase generation to achieve LS. As mentioned previously, in order to satisfy the need to embed secret information in any cover text, we use a self-supervised paraphrase generation method instead of training on a specific paraphrase dataset. Inspired by \cite{ChristianFedermann:2019}, we generate paraphrase by pivoting between two different languages with NMT and select two proximate languages (i.e., English and German) as the pivot languages. We thus name it as an En2Ge2En model.

\begin{figure}[!t]
\centering
\includegraphics[width=3.2in]{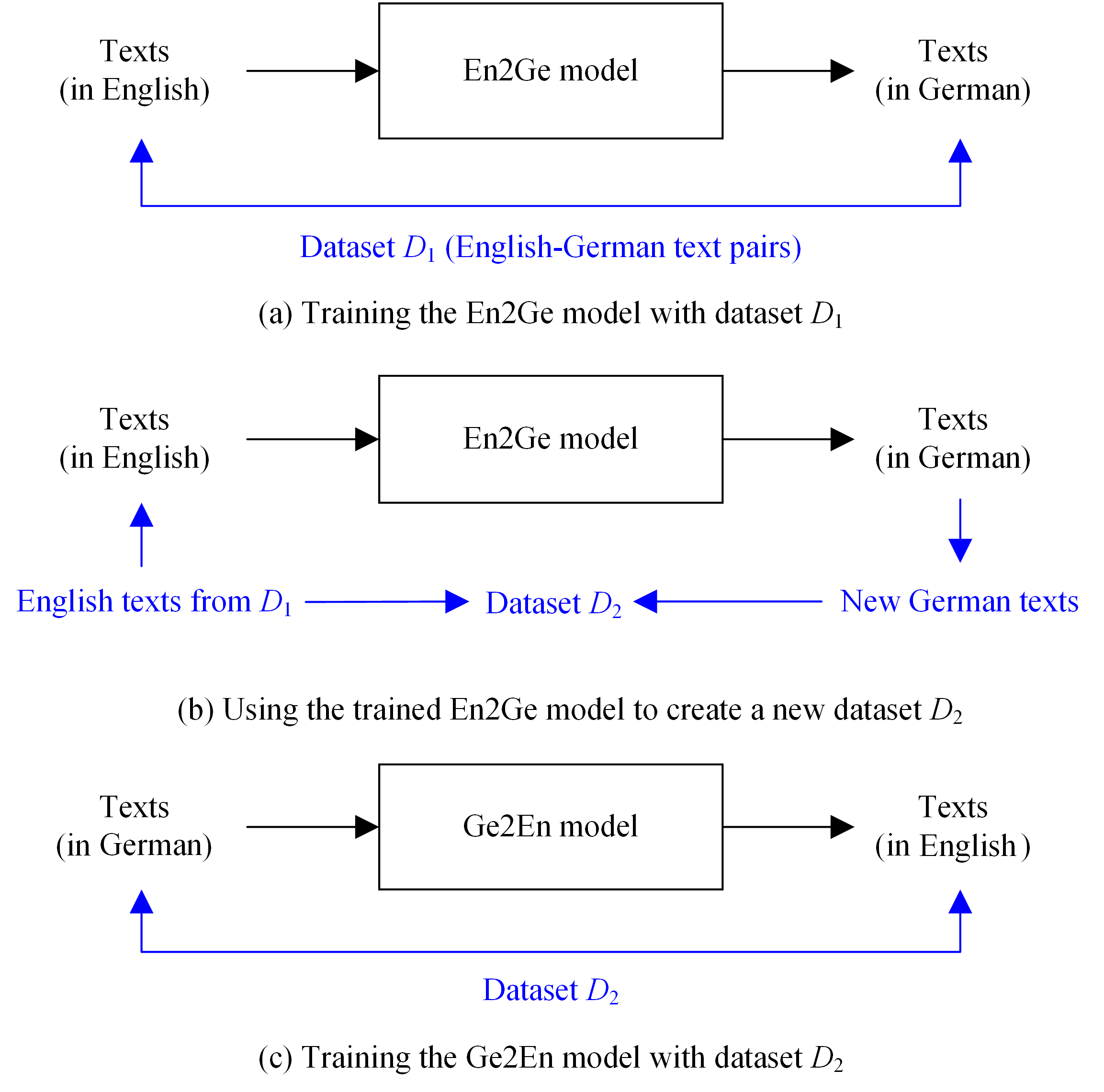}
\caption{Training the En2Ge model and the Ge2En model with two datasets.}
\end{figure}

As has shown in Fig. 1, the En2Ge2En model allows us to translate the cover text in English to another text in German, whose semantic information can be regarded as the same to the cover text. The text in German facilitates us to produce a new text in English, whose semantic information will be quite close to both the cover text and the text in German. Notice that, the resultant text in English generated with the En2Ge2En model will not carry secret information if we do not embed secret information during the language decoding phase shown in Fig. 1. It can be said that, no matter we embed secret information or not, the final text (in English) generated with the En2Ge2En model is expected to have the same meaning to the cover text, and the pivot text in German shown in Fig. 1 can be regarded as the inner semantic meaning of the cover text in English.

A critical problem is how to design the model architecture and its training strategy. Since it is not the main interest of this paper and there are many advanced works in the literature, we use a most popular architecture called Transformer \cite{transformer:paper} as the implementation of the En2Ge2En model. The architecture details can be found in \cite{transformer:paper}. We below describe how to train the En2Ge2En model. It is pointed that training the En2Ge2En model is independent of embedding secret information.

The En2Ge2En model consists of two parts that are English-to-German (En2Ge) and Ge2En. As the basic translation task, we train the En2Ge model with a large corpus containing the corresponding text pairs in English and German. However, when to train the Ge2En model, we create a new dataset. It is due to the reason that the pivot texts (in German) are no longer the original ones in the large corpus when to embed secret data, leading to the distribution difference between texts. Therefore, we use the trained En2Ge model to translate all English texts in the large corpus to generate the corresponding German texts. We then replace the original German texts in the large corpus with the generated German texts and keep the English texts unchanged to form a new dataset. Except for the difference in dataset and input/output languages, we train the Ge2En model in the same way as the En2Ge model. Fig. 2 shows how to train the En2Ge model and Ge2En model with two different datasets. We deem this method self-supervised, whose application scenario is broader compared with previous methods that are constrained by specific dataset(s).

\subsection{Data Embedding}
After training the entire En2Ge2En model, we are to use the trained model to embed secret data into the cover text. Unlike traditional MLS arts that directly modify a few words in the cover text, the proposed work even alters the entire expression of the cover text, which ensures a significantly higher payload.

As mentioned previously, the En2Ge2En model corresponds to a Seq2Seq2Seq model that enables us to output a new text by a word-wise (or called token-wise) fashion. In other words, the En2Ge2En model allows us to orderly generate a sequence of words (or called tokens) and construct a new text by orderly collecting all the words. Without the loss of generalization, let $\textbf{y} = \{y_1, y_2, ..., y_n\}$ represent the newly generated text, where $n > 0$ is the length of the generated text and $y_i$ means the $i$-th word. For consistency, we will also regard $y_i$ as a token. Let $V = \{v_1, v_2, ..., v_m\}$ represent the vocabulary including all tokens, where $m > 0$ is the size of the vocabulary. We can write $y_i \in V$ for all $i\in [1, n]$. The determination of $y_i$ can be briefly summarized as follows. According to the input, the Ge2En model (inside the En2Ge2En model) first outputs the prediction probability for each token in $V$. Formally, for each $y_i$, $i\in [1, n]$, the prediction probability of $v_j \in V$, $j\in [1, m]$ obtained by the Ge2En model is expressed as $p_{i,j} \in [0, 1]$ and
\begin{equation}
\sum_{j=1}^{m}p_{i,j} = 1, \forall i\in [1,n].
\end{equation}

A higher prediction probability implies that the corresponding token is more suited to be selected as the present output. Generally, for each $y_i$, the token with the highest prediction probability can be selected as the output, i.e., 
\begin{equation}
y_i = \underset{v_j\in V}{\text{arg max}}~p_{i,j}.
\end{equation}

Accordingly, a sequence of tokens $\textbf{y} = \{y_1, y_2, ..., y_n\}$ can be generated regardless of the internal processing mechanism of the Ge2En model. Here, we ignore the terms secret data, secret key, side information and information hiding strategy shown in Fig. 1. Moreover, if there are multiple $v_j$ satisfying the objective in Eq. (2),  $y_i$ can be set to any one among them.

Now, in order to embed secret data, we should modify Eq. (2) in such a way that the present output (if it is used for data embedding) not only matches the secret data, but also results in a ``good'' text, i.e., the text $\textbf{y}$ containing secret data should be seemingly normal and semantically close to the cover text $\textbf{c}$ = $\{c_1, c_2, ..., c_{n'}\}$, where $n' > 0$ is the length of the cover text and $c_i$ means the $i$-th token. Mathematically, given $V = \{v_1, v_2, ..., v_m\}$ and the corresponding prediction probabilities $\textbf{p}_i = \{p_{i,1}, p_{i,2}, ..., p_{i, m}\}$ satisfying Eq. (1), if $y_i$ is required to carry a secret stream $\textbf{b}_i \in \{0,1\}^{l_i}$, where $l_i\geq 0$ is the length of the secret stream, the data embedding operation for $y_i$ in this paper is equivalent to determining $y_i$ as:
\begin{equation}
y_i = \underset{v_j\in V,~f(v_j)=\textbf{b}_i}{\text{arg max}}~p_{i,j},
\end{equation}
where $f(v_j)=\textbf{b}_i$ means that the token $v_j$ is mapped to $\textbf{b}_i$ according to $f$, i.e., by setting $y_i = v_j$ if $f(v_j)=\textbf{b}_i$, we can infer that $\textbf{b}_i$ has been embedded into $y_i$. Obviously, we can retrieve $\textbf{b}_i$ from $y_i$ by applying $f^{-1}(y_i) = f^{-1}(v_j) = \textbf{b}_i$. It is noted that $\textbf{p}_i$ can be always obtained by the Ge2En model.

\begin{algorithm}[!t]
 \caption{Pseudocode for the data embedding procedure}
 \begin{algorithmic}[1]
	\renewcommand{\algorithmicrequire}{\textbf{Input:}}
	\renewcommand{\algorithmicensure}{\textbf{Output:}}
	\REQUIRE Cover text $\textbf{c}$, trained En2Ge2En model $\mathcal{M}$, secret data $\textbf{b}$, vocabulary $V$, secret key $\textbf{k}$, mapping function $f$.
	\ENSURE  Stego text $\textbf{y}$.
	\STATE Set a step value $s\geq 1$ according to $\textbf{k}$
	\STATE Empty $\textbf{y}$ and then set $i = t = 1$
	\WHILE {need to generate the next token}
    	\STATE Determine $\textbf{p}_i$ with $\mathcal{M}$, $\textbf{c}$ and $\{y_1, y_2, ..., y_{i-1}\}$
    	\IF {$(t-1)~\text{mod}~s\neq 0$ or $\textbf{b}$ has been embedded}
    	    \STATE Call Eq. (2) to determine $y_i$
    	    \STATE Set $t = t + 1$
    	\ELSE
    	    \STATE Determine $\textbf{b}_i$ based on $\textbf{b}$ and the previously embedded streams (if any)
    	    \STATE Call Eq. (3) to determine $y_i$
    	    \STATE Set $t = 1$
    	\ENDIF
        \STATE Append $y_i$ to $\textbf{y}$ and then set $i = i + 1$
	\ENDWHILE
	\RETURN $\textbf{y}$
 \end{algorithmic}
\end{algorithm}

Therefore, for data embedding, a critical problem is how to design $f$, which maps every token in $V$ to a binary stream. In this paper, we propose a novel coding technique called SaBins to construct $f$, which will be detailed later. Assuming that $f$ has been constructed, the data embedding procedure proposed in this paper is described as \textbf{Algorithm 1}, from which we can find that we use a step $s$ to control the tokens to be embedded. There are two advantages for this design. One is to generalize conventional token-wise data embedding methods for which we have $s = 1$. The other is to maintain semantic consistency between the cover text and the stego text well by using a small $s > 1$. Our experiments will demonstrate the superiority. It is noted that all ``stego tokens'' in \textbf{Algorithm 1} use the same $f$.

\subsection{Data Extraction}
It is straightforward for the data receiver to fully retrieve $\textbf{b}$ from $\textbf{y}$. Clearly, according to the secret key, the data receiver first determines the step value $s$. Then, with $f$, the data receiver can determine all streams carried by the corresponding individual tokens in $\textbf{y}$. By concatenating all the streams, the entire secret data can be reconstructed. 

It is pointed that there is no need for the data receiver to hold the trained En2Ge2En model and the original cover text, which significantly reduces the side information between the data hider and the data receiver compared with many GLS methods that require the data hider and the data receiver to share a trained language model in advance. Moreover, the proposed work enables the data hider to train language models that generate texts of various styles without sharing the models with anyone else, which prevents the language models as privacy data from leaked. One thing to note is that the data hider and the data receiver should share $f$ in advance, which can be constructed by an offline fashion and therefore will not require much computational cost. In addition, since $f$ is only shared between the data hider and the data receiver, it will be very difficult for an attacker to extract secret data from $\textbf{y}$.

\subsection{Semantic-aware Bins (SaBins) Coding}
We propose an efficient method called \emph{SaBins} to determine
\begin{equation}
f: V \mapsto \cup_{i=0}^{\infty }\{0,1\}^i,    
\end{equation}
which maps each token in $V$ to a binary stream. To this end, we divide $V$ to $2^l+1$ disjoint subsets $V_0, V_1, ..., V_{2^l}$, where $l > 0$ is a pre-determined integer. Mathematically, we have
\begin{equation}
V = \cup_{i=0}^{2^l}V_i~\text{and}~V_i\cap V_j = \emptyset,~\forall~0\leq i\neq j\leq 2^l.
\end{equation}

All the tokens in $V_i, 0\leq i\leq 2^l$, are mapped to the same binary stream. In detail, all the tokens in $V_i$, $0\leq i\leq 2^l-1$, are mapped to the same binary stream that corresponds to the index $i$ and have a length of exactly $l$, e.g., all the tokens in $V_3$ are mapped to ``0011'' if we have $l = 4$, i.e., $f(e)$ = ``0011'' if $e\in V_3$ and $l = 4$. However, all the tokens in $V_{2^l}$ are mapped to an empty stream, i.e., all the tokens in $V_{2^l}$ will not carry secret information. Therefore, given the stego text $\textbf{y} = \{y_1, y_2, .., y_n\}$, if all the tokens in $\textbf{y}$ are orderly used to carry secret information, we can retrieve the entire secret data by concatenating $\{f(y_1), f(y_2), ..., f(y_n)\}$, which will be further decoded to the original message according to the secret key. Notice that $f$ should be shared between the data hider and the data receiver in advance. Otherwise, the hidden information cannot be extracted. A critical task is to determine the subsets. We can also consider the subsets as ``bins''. That is why we define this method as a ``bins coding'' method. 

\begin{algorithm}[!t]
 \caption{Pseudocode for the SaBins coding method}
 \begin{algorithmic}[1]
	\renewcommand{\algorithmicrequire}{\textbf{Input:}}
	\renewcommand{\algorithmicensure}{\textbf{Output:}}
	\REQUIRE $V = \{v_1, v_2, ..., v_m\}$, $D_{1,0}$, $l$, a secret key $\textbf{k}'$.
	\ENSURE  $V_0, V_1, ..., V_{2^l}$ (i.e., $f$).
	\STATE Determine $\textbf{h} = \{h(v_1), h(v_2), ..., h(v_m)\}$ based on $D_{1,0}$
	\STATE Determine $\textbf{h}' = \{h(v_{p_1}), h(v_{p_2}), ..., h(v_{p_m})\}$ by sorting $\textbf{h}$ where we have $h(v_{p_1})\geq h(v_{p_2})\geq ...\geq h(v_{p_m})\geq 0$
	\STATE Initialize $V_0 = V_1 = ... = V_{2^l} = \emptyset$
	\STATE Mark all tokens in $V$ as \emph{unprocessed}
	\STATE Set $V_{2^l}$ = \{\textless eos\textgreater\} and then Mark ``\textless eos\textgreater'' as \emph{processed} 
	\FOR {$i = 1, 2, ..., m$}
    	\STATE Determine $C_{p_i}\subset V$
    	\STATE Collect all \emph{unprocessed} tokens in $C_{p_i}$ to build a set $C_{p_i}'$, whose size is expressed as $|C_{p_i}'|$
    	\WHILE {$|C_{p_i}'| > 0$}
    	\STATE Determine $n_s = \text{min}(|C_{p_i}'|, 2^l)$
    	\STATE Randomly select $n_s$ tokens from $C_{p_i}'$ according to $\textbf{k}'$
    	\STATE Randomly select $n_s$ subsets from $\{V_0, V_1, ..., V_{2^l-1}\}$ according to $\textbf{k}'$
    	\STATE Randomly assign the $n_s$ tokens to the $n_s$ subsets so that no two tokens are in the same subset using $\textbf{k}'$
    	\STATE Mark the $n_s$ selected tokens as \emph{processed}
    	\STATE Remove the $n_s$ selected tokens from $C_{p_i}'$
    	\ENDWHILE
	\ENDFOR
	\RETURN $V_0, V_1, ..., V_{2^l}$
 \end{algorithmic}
\end{algorithm}

The proposed SaBins method assigns the most special token ``\textless eos\textgreater'' to $V_{2^l}$ only. Namely, $V_{2^l}$ contains one and exactly one token ``\textless eos\textgreater'', which is generally the last token of each text indicating the end of the text. For example, the sentence ``how are you ? \textless eos\textgreater''  generated by the language model represents the actual sentence ``how are you ?''. Such a special token ``\textless eos\textgreater'' allows us to decide when to stop generating the text. ``\textless eos\textgreater'' will not carry secret information. 

Now, our task is to construct $V_0, V_1, ..., V_{2^l-1}$. A straightforward is assigning each token in $V$ (except for ``\textless eos\textgreater'') to a randomly selected subset from $\{V_0, V_1, ..., V_{2^l-1}\}$, which, however, does not take into account the semantic characteristics and distribution characteristics of tokens and may lead to poor text quality. For example, for tokens whose semantics are close to each other, and the semantics suit the present text generation step best, if they have been assigned to the same subset and do not match the secret stream to be embedded, it forces the text generation model to select an inappropriate token as the present output, which may result in poor quality of the text. It motivates Fang \emph{et al}. \cite{Fang:paper} to explore a variant where a set of common tokens are assigned to all subsets such that the semantic quality of the text is high. However, these common tokens will no longer carry any secret information when they are included in all subsets. These tokens are used much more frequent than other words. As a result, there is a sharp decline for the embedding payload. In order to achieve a better trade-off between text quality and embedding payload, in this paper, expect for ``\textless eos\textgreater'', each of the other tokens in $V$ will be assigned to a subset from $\{V_0, V_1, ..., V_{2^l-1}\}$ based on its distribution characteristics in the dataset $D_1$ in Fig. 2. We use $D_{1,0}$ to denote a set including all English texts in $D_1$. 

Specifically, we first determine the frequency of each token $v_i\in V, 1\leq i\leq m$, appeared in $D_{1,0}$, expressed as $h(v_i)\geq 0$. It is noted that $V$ can be determined by collecting all different tokens in $D_{1,0}$. Then, all tokens are sorted in a descending order according to the frequencies. Without the loss of generalization, Let $\{v_{p_1}, v_{p_2}, ..., v_{p_m}\}$ be the sorted token-sequence, where $h(v_{p_1})\geq h(v_{p_2})\geq ...\geq h(v_{p_m})$ and $\{p_1, p_2, ..., p_m\}$ is a permutation of $\{1, 2, ..., m\}$. Thereafter, for each token $v_{p_i}$ except for ``\textless eos\textgreater'' in the sorted sequence, we determine its substitution set $C_{p_i}$, which means that the corresponding token $v_{p_i}$ can be replaced with any token in $C_{p_i}$ to maintain the semantics of the text. Obviously, the token itself surely belongs to $C_{p_i}$, i.e., $v_{p_i}\in C_{p_i}$. For $C_{p_i}$, we will assign each \emph{unprocessed} token $e\in C_{p_i}$ to a \emph{randomly selected} subset $V_j$. By orderly processing all substitution sets $\{C_{p_1}, C_{p_2}, ..., C_{p_m}\}$, we can finally construct $\{V_0, V_1, ..., V_{2^l-1}\}$. With $\{V_0, V_1, ..., V_{2^l-1}\}$ and $V_{2^l}$, we actually finish the construction of $f$. \textbf{Algorithm 2} shows the detailed pseudocode, from which we can find that the construction of the subsets takes into account the semantic characteristics and the distribution characteristics of tokens. That is why we define it as a \emph{semantic-aware} coding method. 

\emph{Remark 1:} We use the synonym relationship in WordNet\footnote{\url{https://wordnet.princeton.edu/}} to determine $C_{p_i}$ shown in Line 7 in \textbf{Algorithm 2}. WordNet contains around $1.1\times 10^5$ synonym sets. The tokens in each synonym set have the same meaning. A token may appear in multiple synonym sets. For a token $v_{p_i}$, we collect all tokens in synonym sets that contains $v_{p_i}$ to constitute the corresponding substitution set $C_{p_i}$. Though the construction of $C_{p_i}$ can be optimized (since all tokens in $C_{p_i}$ are not semantically related to each other), it is not the main interest of this paper and our experiments show that the above construction has shown good performance. We leave the optimization for our future work.

\emph{Remark 2:} The side information shown in Fig. 1 includes $f$ and all pre-determined parameters such as $s$ and $l$ (which can be controlled by the secret key). The information encoding strategy shown in Fig. 1 has been described in Section III-C.

\emph{Remark 3:} For Line 9 in \textbf{Algorithm 1}, $\textbf{b}_i$ is a stream with a length of $l$, whose determination is based on $\textbf{b}$ and previously embedded streams. For example, if $\textbf{b}$ = ``010111000'' and the previously embedded stream is ``010'', the present stream to be embedded will be ``111''. Here, we actually assume that $l$ divides the length of $\textbf{b}$, which is reasonable for practice since we can always append ``0'' to the end of $\textbf{b}$ to guarantee that $l$ divides the length of $\textbf{b}$. If $\textbf{b}$ cannot be fully carried by a text, multiple texts can be used for LS. 

\begin{table*}[!t]
\caption{Examples due to different parameters. BPW is roughly determined as the ratio between $l$ and $s$ throughout our simulation.}
\centering
\begin{tabularx}{\linewidth}{c|c|X|c|c}
\hline\hline
Cover text & \multicolumn{4}{c}{\emph{In fantastic weather , 214 cyclists came to Illmensee to take on the circuit , over the hills and around the lake .}} \\
\hline
Parameters & BPW & Stego text & BERTScore & PPL\\
\hline
\multirow{2}{*}{$s = \infty, l = 0$} & \multirow{2}{*}{0} & \emph{When the weather was fine , 214 cyclists arrived in Illmensee to take the route over the hills and the lake .} & \multirow{2}{*}{0.9579} & \multirow{2}{*}{1.361}\\
\hline
\multirow{1}{*}{$s = 3, l = 1$} & \multirow{1}{*}{0.33} & \emph{In fine weather 214 cyclists arrived in Illmensee to take the route over the hills and the lake .} & \multirow{1}{*}{0.9571} & \multirow{1}{*}{1.891}\\
\hline
\multirow{2}{*}{$s = 2, l = 1$} & \multirow{2}{*}{0.50} & \emph{In fine weather , we had 214 bike rikers coming to Ill lake , to pick up the route across the hills and the lake .} & \multirow{2}{*}{0.9223} & \multirow{2}{*}{5.167}\\
\hline
\multirow{2}{*}{$s = 3, l = 2$} & \multirow{2}{*}{0.67} & \emph{With fine weather , 214 bicyclists flocked to Lake Illmen to take over the course across the hills and the lake , and the trip was made in the winter .} & \multirow{2}{*}{0.9187} & \multirow{2}{*}{6.478}\\
\hline
\multirow{1}{*}{$s = 1, l = 1$} & \multirow{1}{*}{1.00} & \emph{In fine weather , 2fourteen bike riders arrived to Ill lake , in order for its way across hill and Lake .} & \multirow{1}{*}{0.8852} & \multirow{1}{*}{10.694}\\
\hline
\multirow{2}{*}{$s = 2, l = 2$} & \multirow{2}{*}{1.00} & \emph{With beautiful wear conditions 2fourone bike drivers were arriving inside Illa , to take track all the distance around hills plus See , where you will be surprised by the quality service of all the bikeers .} & \multirow{2}{*}{0.8825} & \multirow{2}{*}{14.854}\\
\hline
\multirow{2}{*}{$s = 3, l = 3$} & \multirow{2}{*}{1.00} & \emph{When the weather became nice , the 214 touring cybermen came into Illmenus lake to assume the route all over the summits and along the lake itself .} & \multirow{2}{*}{0.8971} & \multirow{2}{*}{12.331}\\
\hline
\multirow{2}{*}{$s = 2, l = 3$} & \multirow{2}{*}{1.50} & \emph{When the local weather occurred , 214 local cyriders arrived in Hillside Illa to undertake the distance over top of both the sides of hills and of the river lake to take them .} & \multirow{2}{*}{0.8734} & \multirow{2}{*}{23.008}\\
\hline
\multirow{2}{*}{$s = 1, l = 2$} & \multirow{2}{*}{2.00} & \emph{With beautiful wear conditions 2fourone bike drivers were arriving inside Illa , to take track all the distance around hills plus See , where you will be surprised by the quality service of all the bikeers .} & \multirow{2}{*}{0.8725} & \multirow{2}{*}{53.846}\\
\hline
\multirow{2}{*}{$s = 1, l = 3$} & \multirow{2}{*}{3.00} & \emph{When a fine climate hit 215 cylindarles would visit Easters , as the path crossed either lake Ille and its mountain \#be .} & \multirow{2}{*}{0.8535} & \multirow{2}{*}{173.32}\\
\hline\hline
\end{tabularx}
\end{table*}

\section{Performance Evaluation and Analysis}
\subsection{Dataset and Model}
In the experiments, we train the En2Ge model on the WMT 2016\footnote{\url{https://www.statmt.org/wmt16/}} English-German translation dataset consisting of around $4.5\times 10^6$ sentence-pairs and a vocabulary containing around $3.3\times 10^4$ tokens (or say words) \cite{GoogleNMT:paper}. The official dataset has been split to three disjoint subsets called training set, validation set, and testing set. We randomly choose some sentence-pairs from the training set and move them to the testing set so that the total number of natural English texts in the testing set is exactly equal to 10,000. After training the En2De model with the training set and the validation set, by inputting the original English texts in the two sets to the trained En2De model, we collect all the generated German texts. Thereafter, we train the De2En model according to the original English texts in the two sets and the corresponding generated German texts.

As mentioned in Subsection III-B, we train both the En2Ge model and the Ge2En model with the structure of Transformer, by exploiting the sequence modeling toolkit Fairseq\footnote{\url{https://github.com/pytorch/fairseq}}. The rest of the training details are consistent with \cite{transformer:paper}. To evaluate the translation performance, we decode the translated text with a beam size of 4 and select the best candidate as the result. The BLEU (BiLingual Evaluation Understudy) \cite{bleu:paper} is then used to evaluate the machine-translated texts. Experimental results show that the BLEU scores for En2Ge and Ge2En are 27.83 and 51.10. It implies that the two well-trained models provide satisfactory translation performance. It is pointed that training the two models is independent of data embedding.  
\subsection{Evaluation Metrics}
\subsubsection{Bits per word} We use bits per word (BPW) to measure the amount of embedded information, which is determined as the ratio between the number of embedded bits and the number of tokens excluding ``\textless eos\textgreater'' in the original cover text. We do not count the number of tokens in the steganographic text since even the same cover text will result in different steganographic texts having a different number of tokens because of different parameters. BPW can be controlled by two parameters $s$ and $l$. Generally, BPW will increase when $s$ decreases or $l$ increases. In experiments, $s = \infty$ and $l = 0$ means to choose the token with the highest prediction probability as the present output at each generation step, i.e., there is no embedded information (i.e., BPW = 0), which serves as a baseline for fair comparison. 

\subsubsection{BLEU} BLEU \cite{bleu:paper} is a metric proposed in 2002, commonly used to evaluate the text quality in machine translation tasks. It counts the coincidence frequency of n-gram words between the machine-generated texts and the gold standard. We evaluate the quality of stego texts with BLEU score due to the use of translation models in our method. Obviously, a higher BLEU score means that the stego texts are more similar to the cover texts, i.e., the higher the text quality. However, it should be pointed that our primary task is not translation. 

\subsubsection{BERTScore} BERTScore \cite{BERTScore:paper} is another metric to evaluate the similarity between texts. Because regarding a large number of partial overlaps between texts as semantic similarity is one-sided, BERTScore takes the internal meaning of the texts into consideration. It calculates the similarity according to word vectors between texts. Because the vectors of words with approximate meanings are also similar to each other, BERTScore is able to evaluate in the semantic space.

\subsubsection{Perplexity (PPL)} The above two metrics are designed to evaluate the similarity of the texts, while paying less attention to the fluency. For example, altering the order of words may raise little influence on BLEU and BERTScore, but may cause critical unnaturalness and confuse human's understanding.

Following \cite{Zhou:TDSC:paper}, we evaluate the fluency of our stego texts with PPL, which is defined as the exponential average negative log-likelihood of a token sequence with a pre-trained language model unless otherwise specified. According to its definition, simply speaking, when tokens with high probability are selected during text generation, the resulting PPL will be lower, meaning that the generated text is more fluent.

\subsubsection{Steganalysis Accuracy} Steganalysis evaluates the security of steganographic methods, referring to detecting whether there is secret information from the observed data. In experiments, we assume that the detection part is available to labeled data. We analyze the anti-steganalysis ability of our method under such adverse condition. Steganalysis can be regarded as a classical binary classification problem, that is, the closer the accuracy rate is to 0.5, the harder for the detection part to distinguish between stego texts and natural ones.

Following \cite{Peng:SPL}, we finetune the $\text{BERT}_\text{base-cased}$ model with the default settings in Hugging Face Transformers\footnote{https://huggingface.co/transformers/} as the detection part. For each experiment, we randomly split 10,000 original natural texts and their stego versions, to three disjoint subsets: training set (60\%), validation set (10\%), and testing set (30\%). The discriminators are trained for 60 epochs, with a batch size of 32. The learning rate is $10^{-6}$. Adam \cite{adam:paper} is used as the optimizer. The accuracy is evaluated on the testing set with the model of the highest validation accuracy.

It is noted that the detection accuracy is still much higher than 0.5 even BPW = 0, indicating that the difference between stego texts and cover texts not only comes from the steganographic methods, but also partly from the model itself \cite{SiyuZhang:arXiv}.

\subsection{Qualitative Results}
We first provide some examples to evaluate the stego texts generated by the proposed method. Table I shows the results. It can be observed that different parameters result in different payloads. The stego texts by applying different parameters also have different BERTScores and PPLs, meaning that the quality of the stego texts are different from each other. Specifically, when the payload (i.e., the BPW) increases, the BERTScore will decline and the PPL will increase, indicating that the text quality declines after embedding with a higher payload. This is consistent with many existing steganographic methods. It is inferred that different parameters may correspond to the same payload, e.g., the BPW will equal 1 when $s = l$ (namely, every $s$ tokens will carry $l = s$ bits averagely). For a fixed $l$, when $s$ increases, the text quality will become better. The reason is that by using a higher $s$, secret information can be evenly distributed throughout the entire text, rather than gather in a certain local area, which is conducive to achieving better text quality. In contrast, for a fixed $s$, when $l$ increases, the text quality will become worse because more bits are carried by the stego tokens. Overall, though different parameters result in different performance, the text quality is satisfactory in most cases. In addition, we visualize the statistical distribution of stego texts and natural texts by applying t-SNE \cite{tsne:paper}. It can be seen from Fig. 3 that as BPW decreases, more points of the two colors overlap, implying that the stego texts are more approximate to the natural ones and thus have higher security. To further verify this, we provide more results in the following.

\begin{figure}[!t]
\centering
\includegraphics[width=\linewidth]{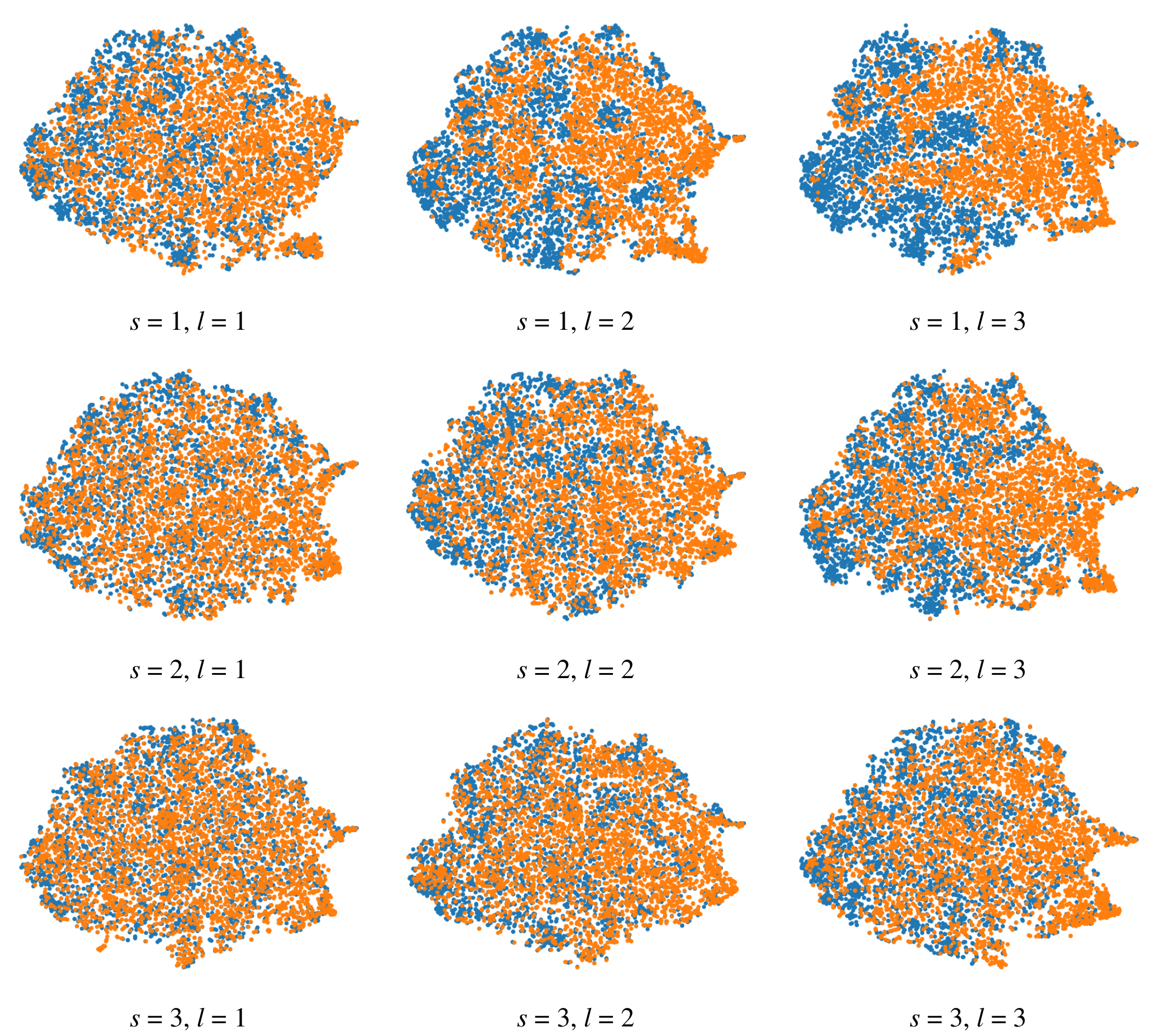}
\caption{Visualization for stego texts and natural ones by applying t-SNE \cite{tsne:paper}. The blue dots represent natural texts and the others are stego texts. }
\end{figure}

\begin{table}
\centering
\caption{Performance comparison between Bins (Baseline) and SaBins (Proposed) due to different parameters.}
\scalebox{0.9}{
\begin{tabular}{c|c|c|c|c|c}
\hline\hline
Parameters & Method & BLEU & BERTScore & PPL & Accuracy \\\hline
$s = \infty,l = 0$ & - & 47.60 & 0.9535 & 1.219 & 0.6422$\pm$0.0183 \\
\hline
\multirow{2}{*}{$s = 3, l = 1$} & Bins & 21.85 & 0.9155 & 2.545 & 0.8351$\pm$0.0076 \\
 & SaBins & \textbf{25.92} & \textbf{0.9213} & \textbf{2.282} & \textbf{0.8111$\pm$0.0047} \\
\hline
\multirow{2}{*}{$s = 2, l = 1$} & Bins & 16.02 & 0.9034 & 3.315 & 0.8638$\pm$0.0090 \\
 & SaBins & \textbf{19.10} & \textbf{0.9083} & \textbf{3.096} & \textbf{0.8458$\pm$0.0043} \\
\hline
\multirow{2}{*}{$s = 3, l = 2$} & Bins & 12.38 & 0.8942 & 3.922 & 0.8915$\pm$0.0137 \\
& SaBins & \textbf{14.87} & \textbf{0.8999} & \textbf{3.616} & \textbf{0.8834$\pm$0.0050} \\
\hline
\multirow{2}{*}{$s = 1, l = 1$} & Bins & 6.86 & 0.8782 & 6.563 & 0.9105$\pm$0.0060 \\
 & SaBins & \textbf{8.58} & \textbf{0.8840} & \textbf{6.019} & \textbf{0.8960$\pm$0.0100} \\
\hline
\multirow{2}{*}{$s = 2, l = 2$} & Bins & 6.70 & 0.8776 & 6.042 & 0.9176$\pm$0.0068 \\
 & SaBins  & \textbf{8.59} & \textbf{0.8835} & \textbf{5.530} & \textbf{0.9092$\pm$0.0038} \\
\hline
\multirow{2}{*}{$s = 3, l = 3$} & Bins & 7.02 & 0.8768 & 5.724 & 0.9167$\pm$0.0061
\\
 & SaBins & \textbf{8.82} & \textbf{0.8799} & \textbf{5.710} & \textbf{0.9086$\pm$0.0055} \\
\hline
\multirow{2}{*}{$s = 2, l = 3$} & Bins & 2.84 & 0.8555 & 9.775 & 0.9522$\pm$0.0037 \\
 & SaBins & \textbf{3.90} & \textbf{0.8620} & \textbf{9.637} & \textbf{0.9485$\pm$0.0038} \\
\hline
\multirow{2}{*}{$s = 1, l = 2$} & Bins & 1.51 & 0.8440 & 16.327 & 0.9778$\pm$0.0043 \\
 & SaBins & \textbf{1.88} & \textbf{0.8491} & \textbf{15.665} & \textbf{0.9560$\pm$0.0030} \\
\hline
\multirow{2}{*}{$s = 1, l = 3$} & Bins & 0.38 & 0.8235 & 33.103 & 0.9977$\pm$0.0047 \\
 & SaBins & \textbf{0.43} & \textbf{0.8252} & \textbf{28.557} & \textbf{0.9786$\pm$0.0036} \\
\hline\hline
\end{tabular}}
\end{table}

\begin{table}
\centering
\caption{Performance comparison between \cite{Ker:NLPStego} and the proposed method at the embedding rate of 0.33 bpw. We use $s = 3, l = 1$ for both methods to ensure that the embedding rates of both methods are equal to each other. PPL is not used here since it is not necessary for traditional MLS.}
\scalebox{0.9}{
\begin{tabular}{c|c|c|c|c}
\hline\hline
Method & Success Rate & BLEU & BERTScore & Accuracy \\
\hline
\cite{Ker:NLPStego} & 93.07\% & \textbf{85.32} & \textbf{0.9753} & 0.8383$\pm$0.0095 \\
Proposed & \textbf{100\%} & 25.92 & 0.9213 & \textbf{0.8111$\pm$0.0047} \\
\hline\hline
\end{tabular}}
\end{table}

\begin{table*}
\centering
\caption{Performance comparison between GLS methods and the proposed method due to different BPWs: 0.33 (low) and 0.67 (high). For those non-embedded tokens in a text (except the pre-fixed tokens), they always have a maximum prediction probability.}
\scalebox{0.9}{
\begin{tabular}{c|c|c|c|c|c|c}
\hline\hline
Method & Parameters & BPW & BLEU & PPL & BERTScore & Accuracy \\
\hline
GPT-2+FLC \cite{Yang:paper}  & \multirow{3}{*}{$s = 3, l = 1$} & \multirow{3}{*}{0.33} & 10.03 & 12.213 & 0.8839 & 0.7610$\pm$0.0208\\
GPT-2+Bins \cite{Fang:paper} &  & & 3.79 & 31.962 & 0.8055 & 0.9996$\pm$0.0012\\
Proposed & & & \textbf{44.07} & \textbf{2.282} & \textbf{0.9426} & \textbf{0.7358$\pm$0.0206}\\
\hline
GPT-2+FLC \cite{Yang:paper} & \multirow{3}{*}{$s = 3, l = 2$} & \multirow{3}{*}{0.67} & 6.45 & 15.419 & 0.8614 & 0.8487$\pm$0.0118\\
GPT-2+Bins \cite{Fang:paper} &  & & 2.45 & 58.133 & 0.8073 & 0.9991$\pm$0.0004\\
Proposed & & & \textbf{22.73} & \textbf{3.096} & \textbf{0.9139} & \textbf{0.8042$\pm$0.0115}\\
\hline\hline
\end{tabular}}
\end{table*}

\begin{table}
\centering
\caption{An example comparison between the word-level strategy and the subword-level strategy. ``\textless unk\textgreater'' means unknown tokens.}
\label{table}
\begin{tabular}{c||c}
\hline\hline
Cover text & There are now 201 cardinals .\\
\hline
German text (with word-level) & Es gibt jetzt 201 \textless unk\textgreater~.\\
\hline
German text (with subword-level) & Es gibt jetzt 201 Kardin$\ddot{\text{a}}$le . \\
\hline
Stego text (with word-level) & There are now \textless unk\textgreater~\textless unk\textgreater~. \\
\hline
Stego text (with subword-level) & There are now the 201 cardinals . \\
\hline\hline
\end{tabular}
\end{table}

\begin{table*}
\centering
\caption{Performance comparison between different strategies. The basic experimental settings are the same as Table IV.}
\scalebox{0.9}{
\begin{tabular}{c|c|c|c|c|c|c}
\hline\hline
Method & Parameters & BPW & BLEU & PPL & BERTScore & Accuracy \\
\hline
GPT-2+Bins \cite{Fang:paper} & \multirow{3}{*}{$s = 3, l = 1$} & 0.33 & 3.79 & 31.962 & 0.8055 & 0.9996$\pm$0.0012\\
GPT-2+Bins+Common-token \cite{Fang:paper} &  & 0.29 & 4.16 & 31.189 & 0.8250 & 0.9797$\pm$0.0016\\
Proposed & & 0.33 & \textbf{44.07} & \textbf{2.282} & \textbf{0.9426} & \textbf{0.7358$\pm$0.0206}\\
\hline
GPT-2+Bins \cite{Fang:paper} & \multirow{3}{*}{$s = 3, l = 2$} & 0.67 & 2.45 & 58.133 & 0.8073 & 0.9991$\pm$0.0004\\
GPT-2+Bins+Common-token \cite{Fang:paper} &  & 0.59 & 3.73 & 50.097 & 0.8122 & 0.9822$\pm$0.0027\\
Proposed & & 0.67 & \textbf{22.73} & \textbf{3.096} & \textbf{0.9139} & \textbf{0.8042$\pm$0.0115}\\
\hline\hline
\end{tabular}}
\end{table*}

\subsection{Performance Comparison with Baseline Method}
We propose a novel method called SaBins for information encoding. Unlike Bins \cite{Fang:paper} that randomly partitions the vocabulary with a fixed size into a certain number of bins evenly in advance, SaBins takes into account the semantic information of tokens and distributes tokens with the similar meaning into different bins so that tokens more fitting into the context can be used for text generation while carrying secret information. In order to verify the superiority of SaBins, we compare SaBins with Bins in terms of different indicators. Table II shows the experimental results. In Table II, for Bins, for example, $l = 2$ means to randomly partition the vocabulary into $2^l = 4$ bins. One more thing to note is that both Bins and SaBins in Table II use pivot translation for fair comparison. In other words, the difference between Bins and SaBins in Table II is that their information encoding strategies are different from each other while the other settings are same as each other. In Table II, for BLEU, BERTScore and PPL, we collect the corresponding average values as the results by default, e.g., each stego text has a PPL and the average value of all PPLs can be determined as the result. For steganalysis, in each experiment, we use ten trained models to determine the detection accuracy. The mean and standard deviation are determined. As shown in Table II, it can be observed that SaBins outperforms Bins in all cases, which has verified the superiority of our work. It also verifies the aforementioned qualitative analysis, e.g., For a fixed $l$, when $s$ increases, the text quality will become better. In addition, the steganalysis performance is dependent on the choice of parameters. It can be inferred from Table II that for a fixed $l$, when $s$ increases, the accuracy will decline, meaning that the security is enhanced. The reason is consistent with the aforementioned qualitative analysis. Similarly, for a fixed $s$, when $l$ increases, the BPW will surely increase, which causes the accuracy to increase and accordingly reduce the security. Overall, by fine-tuning the parameters, a good trade-off between payload, text quality and security can be obtained. 

\subsection{Performance Comparison with Mainstream Methods}
It is necessary to compare the proposed method with MLS since both have the same purpose in maintaining the semantic consistency. Most traditional MLS methods rely on substitute rules or hash mapping, which may lead to a problem that the embedding strategy is unavailable when there are not matching rules or mapping. Table III shows the performance comparison between \cite{Ker:NLPStego} and the proposed method. In \cite{Ker:NLPStego}, Wilson \emph{et al.} exploit both substitute rules with Paraphrase DataBase \cite{PPDB} and hash mapping. The embedding success rate cannot reach 100\%. Alternatively speaking, the success rate also reflects the problem of low embedding payload in MLS methods. It is seen from Table III that the embedding success rate is close to 100\% when the embedding payload is relatively small. However, the proposed method gets rid of such restriction. The stego texts can be generated based on any given cover texts. The method in \cite{Ker:NLPStego} shows satisfactory results in BLEU and BERTScore because of the partial replacement based strategy. However, the result in steganalysis accuracy exposes its unnaturalness and turns out that our method is more ideal in terms of security.

It is also necessary to compare the proposed method with GLS since both embed the secret information in a text generation way. However, GLS methods generate the stego texts without a cover text so that it is difficult to evaluate the quality of the stego text. To solve this problem, we evaluate the quality of the stego text by comparing it with the corresponding zero-bit text, which is generated by the same model but without any hidden information. The zero-bit text is generated by always choosing the token with the largest prediction probability as the present output during text generation. Because RNN-Stega \cite{Yang:paper} and Bins \cite{Fang:paper} do not require any text input, they need to be implemented on a language model. For fair comparison, we finetune another GPT-2 \cite{GPT2:paper} model with the generated English texts mentioned in Subsection IV-A as the language model to conduct experiments. We apply the parameters \emph{s} and \emph{l} to the GLS methods so that we can make comparison under the condition of same BPW. We use the fixed-length coding (FLC) in \cite{Yang:paper} for simulation. And for Bins, the baseline strategy in Table II is used. We use 10,000 stego texts and the corresponding zero-bit texts for each steganalysis experiment. During text generation, we select the tokens with top-10000 appearance frequency in the training set mentioned in Subsection IV-A as the first tokens of the generated stego texts so that no two stego texts are the same as each other, which has been utilized in conventional works. As shown in Table IV, compared to GLS methods, the proposed method better preserves the semantic information in text generation. It turns out that the stego texts generated by the proposed method are more approximate to natural texts, which is also consistent with the requirement of higher security and verified by the steganalysis results.

\begin{figure}[!t]
\centering
\includegraphics[width=\linewidth]{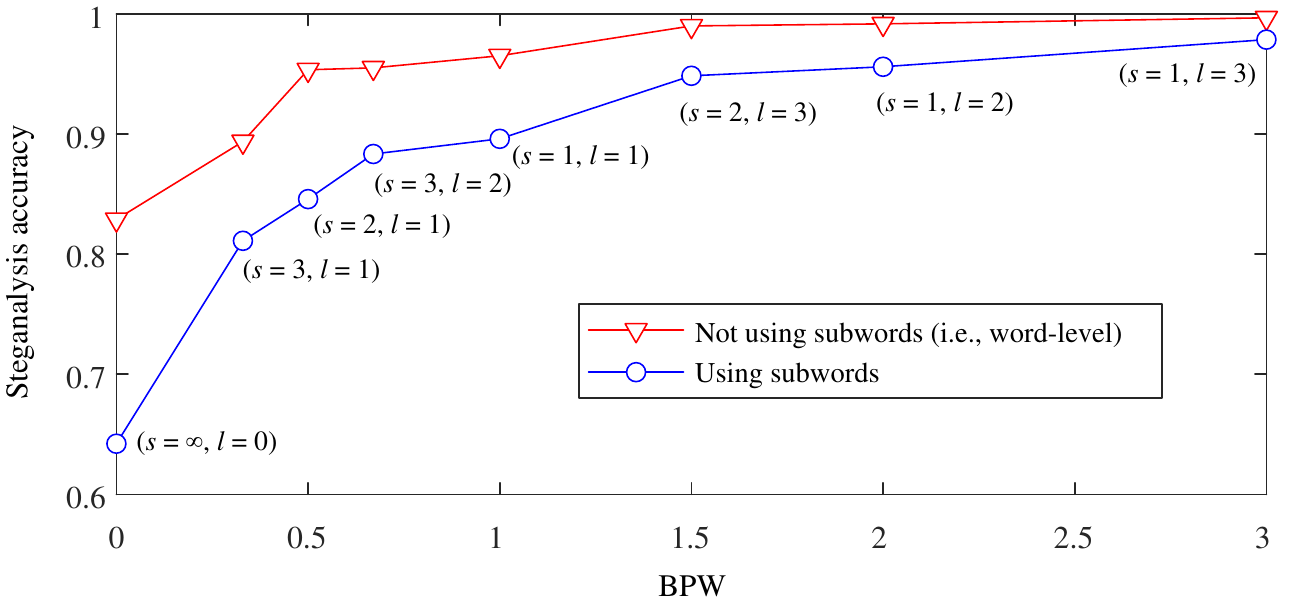}
\caption{Steganalysis accuracy versus BPW for using/not using the subwords.}
\end{figure}

\subsection{Other Comparisons and Complexity Analysis}
Our simulation is based on the subword strategy, which has been mentioned in Subsection II-C. Dividing the words into subwords will reduce the number of synonym relationships in the vocabulary. As a result, in our simulation, the number of synonyms relationship decreases from around $2.6\times 10^4$ to around $9.4\times 10^3$ for the used dataset. However, the text quality shows comparable results with the condition of word-level. That is because, only those words that appear less frequently will be divided into subwords. Due to the low occurrence rate, their influence on SaBins is negligible. However, when implementing the secret embedding strategy on a word-level Seq2Seq2Seq model, the security drops sharply due to a large number of unknown tokens, as shown in Fig. 4. It also confuses human understanding, as shown in Table V, which is more likely to arouse the suspicion of the detector. Generally speaking, embedding in a subword-level is superior to that in a word-level, despite of the reduction of synonym relationships.

The authors in \cite{Fang:paper} propose a novel common-token strategy for improving the quality of the generated stego text. The price of this strategy is that the used common tokens no longer carry secret information, accordingly resulting in a lower payload. In experiments, we select the tokens with top-1000 appearance frequency in the training set mentioned in Subsection IV-A as common tokens and introduce the same parameters in Table IV for fair comparison. As shown in Table VI, although applying common-token improves the text quality sightly compared with ``GPT-2+Bins'', it is still at a low level and more likely detected by the steganalyzer compared with the proposed method, which has demonstrated the superiority of the proposed method.  

In addition, the computational complexity of the proposed method is mainly affected by model training and SaBins. Since both can be completed by an off-line fashion, the proposed method is very suitable for application scenarios based on the aforementioned convinced experiments. 

\section{Conclusion}
LS is not only promising but also challenging. Mainstream MLS methods suffer the problem of low embedding payload, while GLS methods are restricted by semantic inconsistency. In this paper, we propose a novel LS method based on pivot translation to maintain the semantic meaning in an automatic generation manner with a Seq2Seq2Seq model and provide a high payload. We also propose an efficient information encoding method to improve the text quality. Extensive experiments have shown that the proposed method outperforms the baseline method, mainstream MLS methods and GLS methods in terms of various indicators. It is believed that the proposed method has good potential in applications. In future, we will improve our method so that the security can be further enhanced.


\end{document}